\documentclass[showkeys,showpacs,superscriptaddress]{revtex4}
\usepackage{amsmath}
\usepackage{amssymb}
\usepackage{graphicx}
\usepackage{color}

\begin{document}

\title{Dirac star in the presence of Maxwell and Proca fields
}
\author{Vladimir Dzhunushaliev}
\email{v.dzhunushaliev@gmail.com}
\affiliation{
	Institute of Experimental and Theoretical Physics,  Al-Farabi Kazakh National University, Almaty 050040, Kazakhstan
}
\affiliation{
National Nanotechnology Laboratory of Open Type,  Al-Farabi Kazakh National University, Almaty 050040, Kazakhstan
}
\affiliation{
	Department of Theoretical and Nuclear Physics,  Al-Farabi Kazakh National University, Almaty 050040, Kazakhstan
}
\affiliation{
	Institute of Physicotechnical Problems and Material Science of the NAS of the Kyrgyz Republic, 265 a, Chui Street, Bishkek 720071,  Kyrgyzstan
}

\author{Vladimir Folomeev}
\email{vfolomeev@mail.ru}
\affiliation{
	Institute of Experimental and Theoretical Physics,  Al-Farabi Kazakh National University, Almaty 050040, Kazakhstan
}
\affiliation{
National Nanotechnology Laboratory of Open Type,  Al-Farabi Kazakh National University, Almaty 050040, Kazakhstan
}
\affiliation{
	Institute of Physicotechnical Problems and Material Science of the NAS of the Kyrgyz Republic, 265 a, Chui Street, Bishkek 720071,  Kyrgyzstan
}


\begin{abstract}
We consider configurations consisting of a gravitating nonlinear spinor field $\psi$, with
a nonlinearity of the type $\lambda\left(\bar\psi\psi\right)^2$, minimally coupled to Maxwell and Proca fields
through the coupling constants~$Q_M$ [U(1) electric charge] and $Q_P$, respectively.
In order to ensure spherical symmetry of the configurations, we use two spin-$1/2$ fields having opposite spins.
By means of numerical computations, we find families of equilibrium configurations with a positive Arnowitt-Deser-Misner (ADM) mass described by regular zero-node
asymptotically flat solutions for static Maxwell and Proca fields and for stationary spinor fields.
For the case of the Maxwell field, it is shown
that, with increasing charge $Q_M$, the masses of the objects increase and diverge as the charge tends to a critical value.
For negative values of the coupling constant $\lambda$, we demonstrate that,
by choosing physically reasonable values of this constant,
it is possible to obtain configurations with masses comparable to the Chandrasekhar mass
and with effective radii of the order of kilometers.
It enables us to speak of an astrophysical interpretation of such systems, regarding them as charged Dirac stars.
In turn, for the system with the Proca field, it is shown that the mass of the configurations also grows with increasing both $|\lambda|$
and the coupling constant $Q_P$.
Although in this case the numerical calculations do not allow us to make a definite conclusion about the possibility of obtaining
masses comparable to the Chandrasekhar mass for physically reasonable values of $\lambda$,
one may expect that such masses can be obtained for certain values of free parameters of the system under consideration.
\end{abstract}



\maketitle

\section{Introduction}

The bulk of the literature is devoted to treating compact gravitating configurations consisting of various fundamental fields. The most popular line of investigation
focuses on studying boson stars~-- objects supported by scalar (spin-0) fields. Being in their own gravitational field, such fields can create configurations
 whose physical characteristics lie in a very wide range, from those which are typical for atoms up to the parameters comparable with characteristics
 of galaxies~\cite{Schunck:2003kk,Liebling:2012fv}.

On the other hand, it is not impossible that there may exist gravitating objects supported by fundamental fields with nonzero spin.
In particular, they may be massive vector (spin-1) fields described by the Proca equation~\cite{Lawrie2002}.
Being the generalization of Maxwell's theory, Proca theory permits one both to take into account various effects related to the possible presence of the rest mass of a photon~\cite{Tu:2005ge}
and to describe the massive $Z^0$ and $W^\pm$ particles in the Standard Model of particle physics~\cite{Lawrie2002}.
Such fields are also discussed in the literature as applied to dark matter physics~\cite{Pospelov:2008jd} and when considering compact strongly
gravitating spherically symmetric starlike configurations~\cite{Brito:2015pxa,Herdeiro:2017fhv}.

In turn, when the source of gravitation is spinor (spin-$1/2$) fields, the corresponding configurations are described by the Einstein-Dirac equations.
These can be spherically symmetric systems consisting of both linear
 \cite{Finster:1998ws,Herdeiro:2017fhv} and nonlinear spinor fields~\cite{Krechet:2014nda,Adanhounme:2012cm,Dzhunushaliev:2018jhj}.
Nonlinear spinor fields are also used in considering cylindrically symmetric solutions~\cite{Bronnikov:2004uu}, wormhole solutions~\cite{Bronnikov:2009na},
and various cosmological problems (see Refs.~\cite{Ribas:2010zj,Ribas:2016ulz,Saha:2016cbu} and references inside).

The aforementioned localized self-gravitating configurations with spinor fields
are prevented from collapsing under their own gravitational fields due to the Heisenberg uncertainty principle.
If one adds to such systems an electric field, the presence of extra repulsive forces related to such a field results in new effects which can considerably alter
the characteristics of the systems~\cite{Finster:1998ux}. Consistent with this, the purpose of the present paper is to study the influence that the presence of
massless (Maxwell) or massive (Proca) vector fields has on the properties of gravitating configurations consisting of a spinor field
$\psi$ with a nonlinearity of the type $\lambda\left(\bar\psi\psi\right)^2$.
Since the spin of a fermion has an intrinsic orientation in space, a system consisting of a single spinor particle cannot be spherically symmetric.
For this reason, we take two fermions having opposite spins, i.e., consider two spinor fields, and this enables us to have spherically symmetric objects.
In order to get configurations with masses of the order of the Chandrasekhar mass, we study in detail the limiting systems obtained in case of using
large negative values of the dimensionless coupling constant $\bar \lambda$.

Notice here that in the present paper we deal with a system consisting of a {\it classical} spinor field. Following Ref.~\cite{ArmendarizPicon:2003qk},
by the latter we mean a set of four complex-valued spacetime functions that transform according to
the spinor representation of the Lorentz group. But it is evident that realistic spin-$\frac{1}{2}$ particles must be described
by {\it quantum} spinor fields. It is usually believed that there exists no classical limit for quantum spinor fields.
However, classical spinors can be regarded as arising from some effective description of more complex quantum systems (for possible justifications of the existence of classical spinors,
see Ref.~\cite{ArmendarizPicon:2003qk}).

 The paper is organized as follows. In Sec.~\ref{prob_statem}, we present the general-relativistic equations for the systems under consideration.
These equations are solved numerically in Sec.~\ref{num_sol} for the Maxwell field
(Sec.~\ref{Maxw_field}) and for the Proca field (Sec.~\ref{Proca_field}) in two limiting cases when
the coupling constant $\bar \lambda=0$ (linear spinor fields) and when $|\bar\lambda| \gg 1$.
Finally, in Sec.~\ref{concl}, we summarize and discuss the results obtained.

\section{Statement of the problem and general equations}
\label{prob_statem}

We consider compact gravitating configurations consisting of a spinor field minimally coupled to
 Maxwell/Proca fields. The modeling is carried out within the framework of Einstein's general relativity.
The corresponding total action for such a system can be represented in the form
[the metric signature is $(+,-,-,-)$]
\begin{equation}
\label{action_gen}
	S_{\text{tot}} = - \frac{c^3}{16\pi G}\int d^4 x
		\sqrt{-g} R +S_{\text{sp}}+S_{\text{v}},
\end{equation}
where $G$ is the Newtonian gravitational constant;
$R$ is the scalar curvature; and $S_{\text{sp}}$ and $S_{\text{v}}$ denote the actions
of spinor and vector fields, respectively.

The action $S_{\text{sp}}$ is obtained from the Lagrangian for the spinor field  $\psi$ of the mass $\mu$,
\begin{equation}
	L_{\text{sp}} =	\frac{i \hbar c}{2} \left(
			\bar \psi \gamma^\mu \psi_{; \mu} -
			\bar \psi_{; \mu} \gamma^\mu \psi
		\right) - \mu c^2 \bar \psi \psi - F(S),
\label{lagr_sp}
\end{equation}
where the semicolon denotes the covariant derivative defined as
$
\psi_{; \mu} =  [\partial_{ \mu} +1/8\, \omega_{a b \mu}\left( \gamma^a  \gamma^b- \gamma^b  \gamma^a\right)+i Q_{M,P}/(\hbar c) A_\mu]\psi
$.
Here $\gamma^a$ are the Dirac matrices in the standard representation in flat space
 [see, e.g.,  Ref.~\cite{Lawrie2002}, Eq.~(7.27)]. In turn, the Dirac matrices in curved space,
$\gamma^\mu = e_a^{\phantom{a} \mu} \gamma^a$, are obtained using the tetrad
 $ e_a^{\phantom{a} \mu}$, and $\omega_{a b \mu}$ is the spin connection
[for its definition, see Ref.~\cite{Lawrie2002}, Eq.~(7.135)].
The term $i Q_{M,P}/(\hbar c) A_\mu\psi$ describes the interaction between the spinor and Maxwell/Proca fields.
The coupling constant $Q_{M}$ plays the role of a U(1) charge in Maxwell theory, and $Q_P$
is the coupling constant in Proca theory.
This Lagrangian contains an arbitrary nonlinear term $F(S)$, where the invariant $S$ can depend on
$
	\left( \bar\psi \psi \right),
	\left( \bar\psi \gamma^\mu \psi \right)
	\left( \bar\psi \gamma_\mu \psi \right)$, or
	$\left( \bar\psi \gamma^5 \gamma^\mu \psi \right)
	\left( \bar\psi \gamma^5 \gamma_\mu \psi \right)$.

The action for the vector fields $S_{\text{v}}$ appearing in \eqref{action_gen} is obtained from the Lagrangian
$$	L_{\text{v}} =
		-\frac{1}{4} F_{\mu\nu}F^{\mu\nu}+\frac{1}{2}\left(\frac{m_P c}{\hbar}\right)^2 A_\mu A^\mu,
$$
where $F_{\mu\nu} = \partial_{ \mu} A_\nu - \partial_\nu A_\mu$ is the tensor
of a massive spin-1 field of the Proca mass $m_P$. In the case of $m_P=0$ we return to Maxwell's electrodynamics.

Varying the action \eqref{action_gen} with respect to the metric, to the spinor field, and to the vector potential $A_\mu$, we derive the Einstein, Dirac, and Proca/Maxwell equations in curved spacetime:
\begin{eqnarray}
	R_{\mu}^\nu - \frac{1}{2} \delta_{\mu }^\nu R &=&
	\frac{8\pi G}{c^4} T_{\mu }^\nu,
\label{feqs-10} \\
	i \hbar \gamma^\mu \psi_{;\mu} - \mu c \psi - \frac{1}{c}\frac{\partial F}{\partial\bar\psi}&=& 0,
\label{feqs-20}\\
	i \hbar \bar\psi_{;\mu} \gamma^\mu + \mu c \bar\psi +
	\frac{1}{c}\frac{\partial F}{\partial\psi}&=& 0,
\label{feqs-21}\\
\frac{1}{\sqrt{-g}}\frac{\partial}{\partial x^\nu}\left(\sqrt{-g}F^{\mu\nu}\right)&=&-Q_{M,P}\bar\psi\gamma^\mu\psi
+\left(\frac{m_P c}{\hbar}\right)^2 A^\mu.
\label{feqs-22}
\end{eqnarray}
The right-hand side of Eq.~\eqref{feqs-10} contains the energy-momentum tensor
 $T_{\mu}^\nu$, which can be represented (already in a symmetric form) as
\begin{align}
\label{EM}
\begin{split}
	T_{\mu}^\nu =&\frac{i\hbar c }{4}g^{\nu\rho}\left[\bar\psi \gamma_{\mu} \psi_{;\rho}+\bar\psi\gamma_\rho\psi_{;\mu}
-\bar\psi_{;\mu}\gamma_{\rho }\psi-\bar\psi_{;\rho}\gamma_\mu\psi
\right]-\delta_\mu^\nu L_{\text{sp}}
\\
&-F^{\nu\rho}F_{\mu\rho}+\frac{1}{4}\delta_\mu^\nu F_{\alpha\beta}F^{\alpha\beta}+
\left(\frac{m_P c}{\hbar}\right)^2\left(A_\mu A^\nu-\frac{1}{2}\delta_\mu^\nu A_\rho A^\rho\right).
\end{split}
\end{align}
Next, taking into account the Dirac equations \eqref{feqs-20} and \eqref{feqs-21}, the Lagrangian \eqref{lagr_sp} becomes
$$
	L_{\text{sp}} = - F(S) + \frac{1}{2} \left(
		\bar\psi\frac{\partial F}{\partial\bar\psi} +
		\frac{\partial F}{\partial\psi}\psi
	\right).
$$
For our purpose,  we choose the nonlinear term in a simple power-law form,
$F(S) = - k(k+1)^{-1}\lambda\left(\bar\psi\psi\right)^{k+1},
$
where $k$ and $\lambda$ are some free parameters. In what follows we set $k=1$ to give
\begin{equation}
	F(S) = - \frac{\lambda}{2} \left(\bar\psi\psi\right)^2.
\label{nonlin_term_2}
\end{equation}
(Regarding the physical meaning of the constant $\lambda$ appearing here, see below.)
In the absence of gravitation, classical spinor fields with this type of nonlinearity have been considered, for instance,
in Refs.~\cite{Finkelstein:1951zz,Finkelstein:1956,Soler:1970xp},
where it has been shown that the corresponding nonlinear Dirac equation has regular finite energy solutions in a flat spacetime.
In turn, soliton-type solutions of the nonlinear Dirac equation in a curved background have been studied in Ref.~\cite{Mielke:2017nwt}
(see also references therein).

Since here we consider only spherically symmetric configurations, it is convenient to choose the spacetime metric in the form
\begin{equation}
	ds^2 = N(r) \sigma^2(r) (dx^0)^2 - \frac{dr^2}{N(r)} - r^2 \left(
		d \theta^2 + \sin^2 \theta d \varphi^2
	\right),
\label{metric}
\end{equation}
where $N(r)=1-2 G m(r)/(c^2 r)$, and the function $m(r)$ corresponds to the current mass of the configuration
enclosed by a sphere with circumferential radius $r$; $x^0=c t$ is the time coordinate.

In order to describe the spinor field, one must choose the corresponding ansatz for $\psi$
compatible with the spherically symmetric line element \eqref{metric}.
Here, we use a stationary ansatz, which can be taken in the following form (see, e.g., Refs.~\cite{Soler:1970xp,Li:1982gf,Li:1985gf,Herdeiro:2017fhv}):
\begin{equation}
	\psi^T =2\, e^{-i \frac{E t}{\hbar}} \begin{Bmatrix}
		\begin{pmatrix}
			0 \\ - g \\
		\end{pmatrix},
		\begin{pmatrix}
			g \\ 0 \\
		\end{pmatrix},
		\begin{pmatrix}
			i f \sin \theta e^{- i \varphi} \\ - i f \cos \theta \\
		\end{pmatrix},
		\begin{pmatrix}
			- i f \cos \theta \\ - i f \sin \theta e^{i \varphi} \\
		\end{pmatrix}
	\end{Bmatrix},
\label{spinor}
\end{equation}
where $E/\hbar$ is the spinor frequency and
$f(r)$ and $g(r)$ are two real functions.
This ansatz ensures that the spacetime of the system under consideration remains static. Here, each row
 describes a spin-$\frac{1}{2}$ fermion, and these two fermions have the same masses $\mu$ and opposite spins.
Thus the ansatz  \eqref{spinor} describes two Dirac fields whose
energy-momentum tensors are not spherically symmetric, but their sum gives a spherically symmetric energy-momentum tensor.
(Regarding the relationships between the {\it Ansatz} \eqref{spinor} and {\it Ans\"{a}tze} used in the literature, see Ref.~\cite{Dzhunushaliev:2018jhj}.)

For the Maxwell and Proca fields, we take the {\it Ansatz} $A_\mu=\{\phi(r),0,0,0\}$.
In the case of Maxwell's electrodynamics this corresponds to the presence of the radial electric field $E_r=-\phi^\prime(r)$
Then, substituting the {\it Ansatz} \eqref{spinor} and the metric  \eqref{metric} into the field equations
\eqref{feqs-10}, \eqref{feqs-20}, and  \eqref{feqs-22}, one can obtain the following set of equations:
\begin{eqnarray}
	&&\bar f^\prime + \left[
		\frac{N^\prime}{4 N} + \frac{\sigma^\prime}{2\sigma}+\frac{1}{x}\left(1+\frac{1}{\sqrt{N}}\right)
	\right] \bar f + \left[
		\frac{1}{\sqrt{N}} +8\bar \lambda\,\frac{\bar f^2 - \bar g^2}{\sqrt{N}}- \frac{1}{\sigma N} \left(\bar E-\bar Q_{M,P} \bar\phi\right)
	\right]\bar g= 0,
\label{fieldeqs-1_dmls}\\
	&&\bar g^\prime + \left[
			\frac{N^\prime}{4 N} + \frac{\sigma^\prime}{2\sigma} +
			\frac{1}{x}\left(1 - \frac{1}{\sqrt{N}}\right)
	\right]\bar g + \left[
		\frac{1}{\sqrt{N}} + 8\bar \lambda\,\frac{\bar f^2 - \bar g^2}{\sqrt{N}}+\frac{1}{\sigma N} \left(\bar E-\bar Q_{M,P} \bar\phi\right)
	\right]\bar f= 0,
\label{fieldeqs-2_dmls}\\
	 &&\bar m^\prime=8 x^2\left[
\frac{\bar f^2+\bar g^2}{\sigma\sqrt{N}}\left(\bar E-\bar Q_{M,P} \bar\phi\right)+4\bar \lambda\left(\bar f^2-\bar g^2\right)^2
+\frac{1}{16}\frac{\bar\phi^{\prime 2}}{\sigma^2}+\frac{\alpha^2}{16}\frac{\bar\phi^{2}}{N\sigma^2}
\right],
\label{fieldeqs-3_dmls}\\
&&\frac{\sigma^\prime}{\sigma}	=\frac{8 x}{\sqrt{N}}\left[
\frac{\bar f^2+\bar g^2}{\sigma N}\left(\bar E-\bar Q_{M,P} \bar\phi\right)+ \bar g \bar f^\prime-\bar f \bar g^\prime+\frac{\alpha^2}{8}\frac{\bar\phi^{2}}{N^{3/2}\sigma^2}
\right],
\label{fieldeqs-4_dmls}\\
&&\bar \phi^{\prime\prime}+\left(\frac{2}{x}-\frac{\sigma^\prime}{\sigma}\right)\bar\phi^\prime=-8\bar Q_{M,P} \frac{\sigma}{\sqrt{N}}\left(\bar f^2+\bar g^2\right)
+\alpha^2\frac{\bar\phi}{N},
\label{fieldeqs-5_dmls}
\end{eqnarray}
where the prime denotes differentiation with respect to the radial coordinate.
Here,  Eqs.~\eqref{fieldeqs-3_dmls} and \eqref{fieldeqs-4_dmls} are the  ($^0_0$) and  $[(^0_0)~-~(^1_1)]$ components of the Einstein equations,
respectively. The above equations are written in terms of the following dimensionless variables and parameters:
\begin{eqnarray}
\begin{split}
\label{dmls_var}
	x = & r/\lambda_c, \quad
	\bar E = \frac{E}{\mu c^2}, \quad
	\bar f, \bar g = \sqrt{4\pi}\lambda_c^{3/2}\frac{\mu}{M_\text{Pl}} f, g,\quad
	\bar m = \frac{\mu}{M_\text{Pl}^2} m, \quad
  \bar \phi = \frac{\sqrt{4\pi G}}{c^2}\phi,
\\
	\bar \lambda = & \frac{1}{4\pi \lambda_c^3\mu c^2}
	\left( M_\text{Pl}/\mu\right)^2\lambda,  \quad
  \bar Q_{M,P} = \frac{1}{\sqrt{4\pi G}}\frac{Q_{M,P}}{\mu}, \quad
  \alpha = \frac{m_P}{\mu},
\end{split}
\end{eqnarray}
 where $M_\text{Pl}$ is the Planck mass and $\lambda_c=\hbar/\mu c$ is the constant having
the dimensions of length (since we consider a classical theory, $\lambda_c$ need not be associated with the Compton length);
 the metric function $N=1-2\bar m/x$. Notice here that, using the Dirac equations \eqref{fieldeqs-1_dmls} and \eqref{fieldeqs-2_dmls},
 one can eliminate the derivatives of $\bar f$ and $\bar g$ from the right-hand side of Eq.~\eqref{fieldeqs-4_dmls}.

We conclude this section with the expression for the effective radial pressure $p_r=-T_1^1$ [the $\left(_1^1\right)$ component of the energy-momentum tensor \eqref{EM}].
Using the Dirac equations \eqref{fieldeqs-1_dmls} and
\eqref{fieldeqs-2_dmls}, it can be represented in the following dimensionless form:
$$
\bar p_r\equiv\frac{p_r}{\gamma} = 8\left[
\frac{1}{\sigma\sqrt{N}} \left(\bar E-\bar Q_{M,P} \bar \phi \right)
\left(\bar f^2 + \bar g^2\right)+\left(\bar f^2 - \bar g^2\right)-2\frac{\bar f\bar g}{x}+4\bar \lambda\left(\bar f^2 - \bar g^2\right)^2
\right]-\frac{1}{2}\frac{\bar\phi^{\prime 2}}{\sigma^2}+\frac{\alpha^2}{2}\frac{\bar\phi^{2}}{N\sigma^2},
$$
where $\gamma=c^2 M_{\text{Pl}}^2/\left(4\pi \mu\lambda_c^3\right)$.
This expression permits us to see the physical meaning of the coupling constant $\bar\lambda$: the case of $\bar\lambda > 0$ corresponds to the attraction
and the case of $\bar\lambda < 0$ to the repulsion.

\section{Numerical results}
\label{num_sol}

In performing numerical integration of Eqs.~\eqref{fieldeqs-1_dmls}-\eqref{fieldeqs-5_dmls},
we start from the center of the configuration where some values of the spinor field, $\bar g_c$, of the scalar potential, $\bar \phi_c$, and of the metric function, $\sigma_c$, are given.
The boundary conditions in the vicinity of the center are taken in the form
\begin{equation}
	\bar g\approx \bar g_c + \frac{1}{2}\bar g_2 x^2, \quad
	\bar f\approx \bar f_1 x,
	\quad \sigma\approx \sigma_c+\frac{1}{2}\sigma_2 x^2, \quad
	\bar m\approx \frac{1}{6}\bar m_3 x^3, \quad \bar\phi \approx \bar \phi_c + \frac{1}{2}\bar \phi_2 x^2.
\label{bound_cond}
\end{equation}
Expressions for the expansion coefficients $\bar f_1,  \bar m_3, \sigma_2, \bar g_2, \bar \phi_2$ can be found from
Eqs.~\eqref{fieldeqs-1_dmls}-\eqref{fieldeqs-5_dmls}. In turn, the expansion coefficients
$\sigma_c$, $\bar g_c$, $\bar \phi_c$, and also the parameter $\bar E$,
are arbitrary. Their values are chosen so as to obtain regular and asymptotically flat solutions when
the functions $N(x\to \infty),\sigma(x\to \infty) \to 1$, and $\bar\phi(x\to \infty) \to 0$.
In this case, the asymptotic value of the function
$\bar m(x\to \infty) \equiv \bar m_\infty$ will correspond to the Arnowitt-Deser-Misner (ADM) mass of the configurations under consideration.
Notice here that in the present paper we consider only configurations described by zero-node solutions.

Since the spinor fields decrease exponentially with distance as $\bar g, \bar f \sim e^{-\sqrt{1-\bar E^2}\,x}$ [see Eq.~\eqref{asymp_Proca} below],
numerical calculations are performed up to some boundary point $x_b$
where the functions $\bar g, \bar f $, and their derivatives go to zero (let us refer to such solutions as interior ones).
The location of the boundary point is determined both by the central values $\bar g_c, \bar \phi_c, \sigma_c$
and by the magnitudes of the parameters $\bar Q_{M,P}, \bar \lambda$, and  $\alpha$. The interior solutions are matched with the exterior ones for
 $\bar \phi$, for the mass function $\bar m$, and (in the case of the Proca field) for the metric function  $\sigma$ on the boundary $x=x_b$ (see below).

\subsection{The case of the Maxwell field}
\label{Maxw_field}

\subsubsection{Linear spinor fields}
\label{num_sol_Max_linear}

\begin{figure}[t]
		\begin{center}
			\includegraphics[width=.5\linewidth]{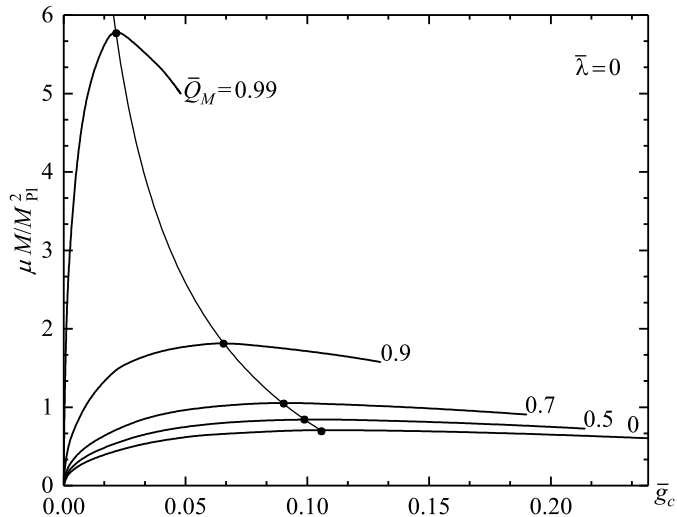}
		\end{center}
\vspace{-0.5cm}
		\caption{Maxwell field: the dimensionless total mass $\bar m_{\infty}$ as a function of  $\bar g_c$ for the systems with $\bar \lambda=0$ and
$\bar Q_M=0, \,0.5,\, 0.7,\, 0.9, 0.99$. The bold dots show the positions of maxima of the mass.
		}
		\label{fig_mass_gc_Q_Maxw}
\end{figure}

 Consider first the case of linear spinor fields, i.e., the problem with $\bar \lambda =0$~\cite{Finster:1998ux}.
 In this case, depending on the values of $\bar g_c$ and $\bar Q_M$,
 the value of $x_b$ is of the order of several hundreds for $\bar g_c \approx 0$,
and it decreases down to $x_b\sim 10$ with increasing $\bar g_c$.
While $\bar g_c$ increases, the mass of the configurations also increases, reaching its maximum whose magnitude depends on the value of $\bar Q_M$.
When $\bar Q_M=0$, the maximum mass is $M^{\text{max}}_{\bar Q_M=0}\approx 0.709 M_\text{Pl}^2/\mu$
 \cite{Herdeiro:2017fhv}. The inclusion of the electric field results in increasing the maximum mass, as is illustrated in Fig.~\ref{fig_mass_gc_Q_Maxw}.

 The presence of the long-range electric field leads to the fact that, besides the interior solutions for the spinor and electric fields, there is also an exterior solution for the electric field.
  To obtain it, we employ Eqs.~\eqref{fieldeqs-3_dmls} and \eqref{fieldeqs-5_dmls}
 by setting $\bar f, \bar g=0$ in them. Also, since for $x=x_b$ the metric function $\sigma=\text{const.}$
 [see Eq.~\eqref{fieldeqs-4_dmls}], we normalize its value to 1 at this point; i.e., we set $\sigma(x_b)=1$.
 [This can always be done since Eqs.~\eqref{fieldeqs-1_dmls}-\eqref{fieldeqs-5_dmls} are invariant under the replacements $\bar \phi \to a \bar\phi,  \sigma \to a \sigma, \bar E \to a \bar E$,
  where $a$ is an arbitrary constant.] As a result, we have the following equations valid for $x>x_b$:
 $$
 \bar \phi^{\prime\prime}+\frac{2}{x}\bar \phi=0, \quad \bar m^\prime=\frac{x^2}{2}\bar \phi^{\prime 2}.
 $$
 As boundary conditions for these equations, we take the corresponding values of the functions $\bar m(x_b)$, $\bar \phi(x_b)$, and $\bar \phi^\prime(x_b)$ on the boundary $x=x_b$.
 Then the exterior solutions are
\begin{equation}
  \bar \phi=C_2+\frac{C_1}{x}, \quad \bar m=\bar m_\infty-\frac{1}{2}\frac{C_1^2}{x},
  \label{asymp_sol}
\end{equation}
where the integration constants are $C_1=-\bar \phi^\prime(x_b) x_b^2, C_2=\bar \phi(x_b)+\bar \phi^\prime(x_b) x_b$.
 In view of the gauge invariance of the Maxwellian electromagnetic field, one can always add to the scalar potential $\bar \phi$ an arbitrary constant $\bar\phi_\infty$
 which ensures that $\bar \phi(\infty)=0$ (this in turn assumes that the constant  $C_2$ is zero).
For Eqs. \eqref{fieldeqs-1_dmls}-\eqref{fieldeqs-5_dmls}, this gauge transformation looks like
  $\bar E-\bar Q_M \bar\phi\equiv \left(\bar E-\bar Q_M \bar\phi_\infty \right)-\bar Q_M\left(\bar\phi-\bar\phi_\infty \right)$.
  Then the only free parameter is $\bar E$, and the solution of the problem reduces to determining such eigenvalues of $\bar E$ for which regular monotonically damped solutions do exist.
  In this case the contribution of the external electric field to the total mass $\bar m_\infty$ is given by the term $C_1^2/(2 x_b)$, and the external metric
    $
  N=1-2 \bar m_\infty/x+C_1^2/x^2
  $
  corresponds to the  Reissner-Nordstr\"{o}m metric.

The total ADM mass of the system [given by the constant  $\bar m_\infty$ from Eq.~\eqref{asymp_sol}] is shown in Fig.~\ref{fig_mass_gc_Q_Maxw}
as a function of $\bar g_c$. In plotting these dependencies, we have kept track of the sign of the binding energy (BE),
which is defined as the difference between the energy of $N_f$ free particles, ${\cal E}_f=N_f \mu c^2$, and the total energy of the system, ${\cal E}_t=M c^2$,
i.e., $\text{BE}={\cal E}_f-{\cal E}_t$. Here, the total particle number
 $N_f$ (the Noether charge) is calculated using the timelike component of the 4-current $j^\alpha=\sqrt{-g}\bar \psi \gamma^\alpha \psi$
as
$
N_f=\int j^t d^3 x,
$
where in our case $j^t = N^{-1/2}r^2 \sin{\theta} \left(\psi^\dag \psi\right)$.
In the dimensionless variables \eqref{dmls_var}, we then have
\begin{equation}
N_f=8\left(\frac{M_\text{Pl}}{\mu}\right)^2\int_0^\infty \frac{\bar f^2+\bar g^2}{\sqrt{N}}x^2 dx.
 \label{part_num}
\end{equation}
A necessary condition for the energy stability is the positiveness of the binding energy.
Therefore, since configurations with a negative BE are certainly unstable, the graphs in Fig.~\ref{fig_mass_gc_Q_Maxw}
are plotted only up to $\bar g_c$ for which the BE becomes equal to 0 (the very right points of the curves).

It is seen from Fig.~\ref{fig_mass_gc_Q_Maxw} that the inclusion of the electric field does not change the qualitative behavior of the curve mass--central density of the spinor field.
With increasing charge, the location of the maximum moves towards smaller values of $\bar g_c$, and the mass increases simultaneously.
The positions of maxima of the mass $M^{\text{max}}$
are joined by a separate curve. Its behavior indicates that while approaching a critical charge, $\bar Q_M \to \bar Q_{\text{crit}}=1$, the quantity
 $\bar g_c \to 0$, and the mass diverges. The asymptotic behavior of this curve for large $\bar Q_M$ can be approximated by the following expression:
  $M^{\text{max}}\approx 0.58 \left(\bar Q_{\text{crit}}-\bar Q_M\right)^{-1/2} M_\text{Pl}^2/\mu$.

When $\bar Q_M > \bar Q_{\text{crit}}$, the behavior of the solutions alters drastically: there are no longer monotonically damped solutions, but there only exist solutions describing
asymptotically damped oscillations. Such solutions correspond to systems whose current mass $\bar m(x)$ diverges as $x\to \infty$,
and, in turn, the spacetime is not asymptotically flat.
The reason for such a change in the behavior of the solutions is intuitively understood from considering the nonrelativistic limit.
 Namely, according to Newton's and Coulomb's laws, the force between two charged, massive point particles is
  $$
 F=-\frac{G \mu^2}{r^2}+\frac{1}{4\pi}\frac{Q_M^2}{r^2}=\frac{G\mu^2}{\lambda_c^2}\frac{\bar Q_M^2-1}{x^2}.
 $$
 Correspondingly, when $\bar Q_M<1$, the gravitational attraction is larger than the electrostatic repulsion, and this enables obtaining gravitationally closed systems.
 In turn, when $\bar Q_M>1$, the repulsive forces dominate, and this leads to destroying the system.

The behavior of the curves shown in Fig.~\ref{fig_mass_gc_Q_Maxw} is similar to the behavior of the corresponding mass--central density dependencies for
boson stars supported by complex scalar fields
(see, e.g., Refs.~\cite{Colpi:1986ye,Gleiser:1988ih,Herdeiro:2017fhv}). In the case of boson stars, the stability analysis against linear perturbations indicates that
the first peak in the mass corresponds to the point separating stable and unstable configurations~\cite{Gleiser:1988ih,Jetzer:1989us}.
One might expect that for the Dirac stars under consideration a similar situation will occur. But this question requires special study.

\subsubsection{Limiting configurations for $|\bar \lambda| \gg 1$}
\label{lim_conf_M}

The main purpose of the present paper is to study the effects of the inclusion of Maxwell and Proca vector fields in the systems consisting of spinor fields
with a nonlinearity of the type~\eqref{nonlin_term_2}. In the absence of vector fields, in our recent paper~\cite{Dzhunushaliev:2018jhj},
it was shown that in the limiting case of large negative values of  $|\bar \lambda| \gg 1$ it is possible to obtain solutions describing configurations with  masses
comparable to the Chandrasekhar mass. In this paper, we extend that problem by including the Maxwell and Proca fields minimally coupled to the spinor fields.

It was demonstrated in Ref.~\cite{Dzhunushaliev:2018jhj} that, with increasing $|\bar \lambda|$,
the maximum total mass of the configurations increases as
 $	M^{\text{max}}\approx \beta \sqrt{|\bar \lambda|}M_\text{Pl}^2/\mu$. The numerical calculations indicate that the inclusion of the charge
$\bar Q_M$ does not lead to qualitative changes, and in the limit $|\bar \lambda| \gg 1$ the above dependence remains the same, and the numerical value of the coefficient
$\beta$ is determined by the magnitude of the charge.

\begin{figure}[t]
	\begin{minipage}[t]{.49\linewidth}
		\begin{center}
			\includegraphics[width=1\linewidth]{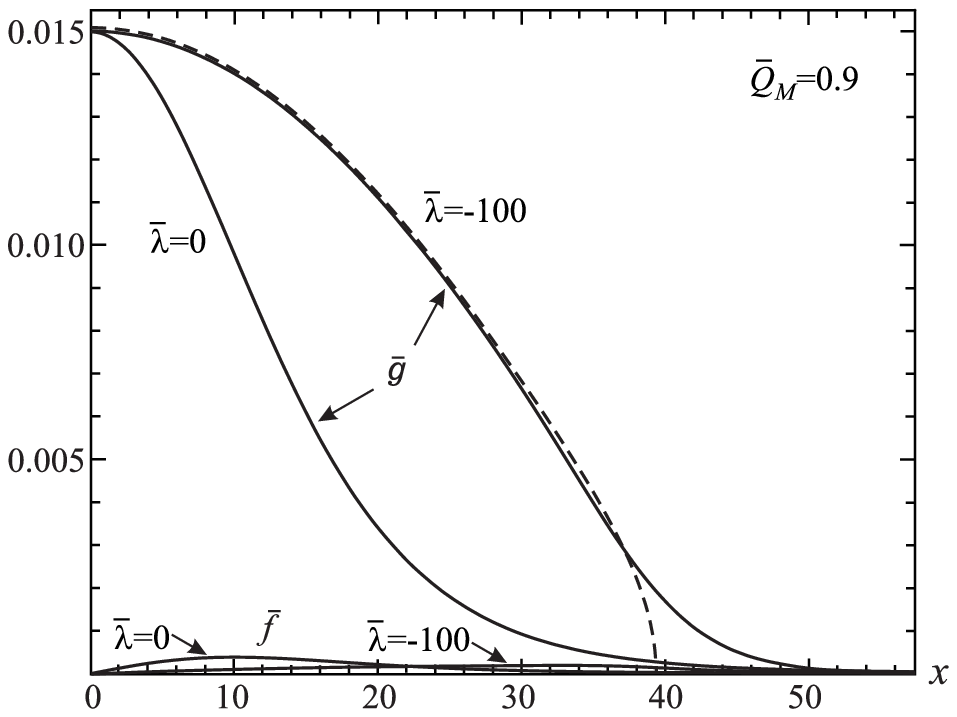}
		\end{center}
\vspace{-0.5cm}
		\caption{Maxwell field: the spinor fields
		 $\bar g$ and $\bar f$  as functions of the dimensionless radius $x$ for $\bar \lambda=-100$ and $\bar \lambda=0$.
The dashed line shows the solution to the approximate equations~\eqref{g_approx}-\eqref{fieldeqs-5_dmls_approx}  with $\bar E/\sigma_c$
from the exact  $\bar g_c=0.015, \bar \lambda=-100$ model, scaled to $\bar \lambda=-100$.
				}
		\label{fig_field_distr}
	\end{minipage}\hfill
\begin{minipage}[t]{.49\linewidth}
		\begin{center}
			\includegraphics[width=1\linewidth]{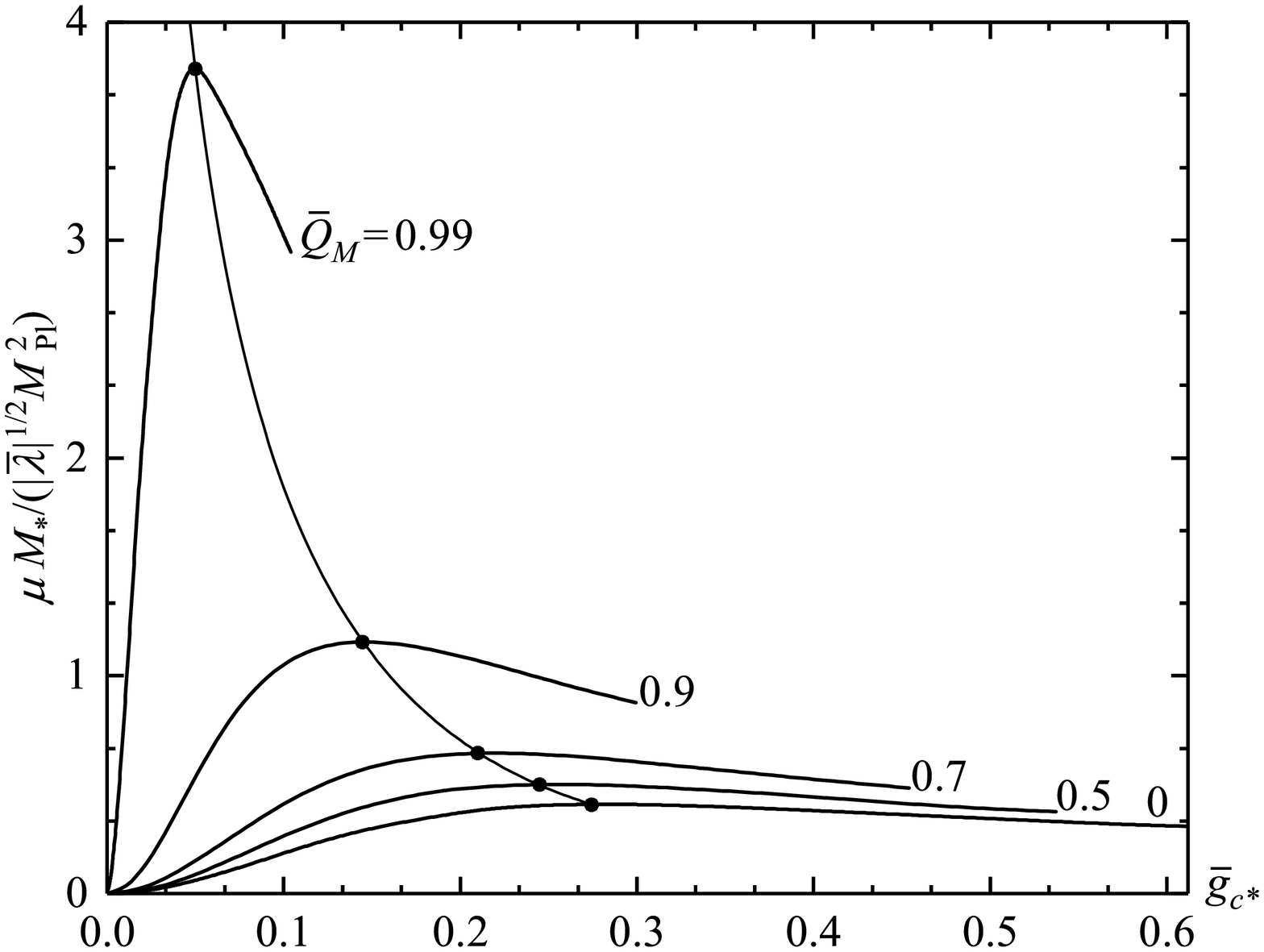}
		\end{center}
\vspace{-0.5cm}
		\caption{Maxwell field: the dimensionless total mass $\bar m_{*\infty}$ as a function of
 $\bar g_{c*}$ for the limiting configurations described by Eqs.~\eqref{g_approx}-\eqref{fieldeqs-5_dmls_approx}.
 The numbers near the curves correspond to the values of the charge $\bar Q_M$.
 The graphs are plotted only for the values of $\bar g_{c*}$ for which the binding energy is positive.
 The bold dots show the positions of maxima of the mass.
  		}
		\label{fig_mass_gc_approx_Maxw}
	\end{minipage}
\end{figure}

In order to obtain the dependence $\beta(\bar Q_M)$,
let us consider an approximate solution to the set of equations~\eqref{fieldeqs-1_dmls}-\eqref{fieldeqs-5_dmls} in the limit
$|\bar \lambda| \gg 1$. To do this, as in the case of uncharged systems of Ref.~\cite{Dzhunushaliev:2018jhj},
one can introduce the following alternative nondimensionalization caused by the scale invariance of these equations: $\bar g_*,
\bar f_*=|\bar \lambda|^{1/2}\bar g,
\bar f, \bar m_*=|\bar \lambda|^{-1/2}\bar m$, and $x_*=|\bar \lambda|^{-1/2}x$.
Using these new variables and taking into account the results of numerical calculations, according to which the leading term
in Eq.~\eqref{fieldeqs-1_dmls} is the third term $(\ldots)\bar g$ and $\bar f$ is much smaller than $\bar g$, this equation yields
\begin{equation}
 \label{g_approx}
 \bar g_* = \sqrt{-\frac{1}{8}
 \left[1 - \frac{1}{\sigma\sqrt{N}}\left(\bar E-\bar Q_M\bar\phi\right)\right]}.
\end{equation}
Substituting this expression into Eqs.~\eqref{fieldeqs-3_dmls}-\eqref{fieldeqs-5_dmls}, one has (again in the approximation of $\bar f \ll \bar g$)
\begin{eqnarray}
	\frac{d \bar m_*}{d x_*} &=& 8 x_*^2 \Big\{
\left[\frac{1}{\sigma\sqrt{N}} \left(\bar E-\bar Q_M\bar\phi\right)- 4\bar g_*^2
\right]\bar g_*^2+\frac{1}{16}\frac{\bar\phi^{\prime^2}}{\sigma^2}		
	\Big\},
\label{fieldeqs-3_dmls_approx}\\
	\frac{d \sigma}{d x_*}& =& \frac{8 x_*}{N^{3/2}}\left(\bar E-\bar Q_M\bar\phi\right)\bar g_*^2,
\label{fieldeqs-4_dmls_approx}\\
\frac{d^2\bar \phi}{d x_*^2}+\left(\frac{2}{x_*}-\frac{1}{\sigma}\frac{d\sigma}{d x_*}\right)\frac{d\bar\phi}{d x_*}&=&-8\bar Q_M \frac{\sigma}{\sqrt{N}}\bar g_*^2,
\label{fieldeqs-5_dmls_approx}
\end{eqnarray}
where now $N=1-2\bar m_*/x_*$. As  $|\bar \lambda|$ increases, the accuracy of
Eqs.~\eqref{g_approx}-\eqref{fieldeqs-5_dmls_approx} becomes better.
This is illustrated in Fig.~\ref{fig_field_distr} where the results of calculations for the configurations with the same
$\bar g_{c}$ and for $\bar \lambda=0$ and  $-100$ are shown.
From comparison of the exact and approximate solutions, one can see their good agreement for the case of $\bar \lambda=-100$,
except the behavior at large radii.
As  $|\bar \lambda|\to \infty$, this region becomes less important and, accordingly, the mass of the configurations will be well described by the asymptotic formula
(see also the discussion of this question in Ref.~\cite{Dzhunushaliev:2018jhj}).

Since $\bar \lambda$ does not appear explicitly in Eqs.~\eqref{fieldeqs-3_dmls_approx}-\eqref{fieldeqs-5_dmls_approx},
one can use these limiting equations to determine the rescaled total mass
$\bar m_{*\infty}\equiv\bar m_*(x\to\infty) = M_{*}/\left(|\bar \lambda|^{1/2}M_\text{Pl}^2/\mu\right)$
as a function of the central density of the spinor field $\bar g_{c*}$.
The corresponding results of a numerical solution to Eqs.~\eqref{g_approx}-\eqref{fieldeqs-5_dmls_approx} are given in Fig.~\ref{fig_mass_gc_approx_Maxw}.
The effect of the inclusion of the charge is similar to the case with $\bar \lambda=0$ from Sec.~\ref{num_sol_Max_linear}:
maximum values of the mass grow with increasing $\bar Q_M$, and the locations of the maxima move towards smaller values of $\bar g_{c*}$.

It follows from the results of solving the approximate set of equations~\eqref{g_approx}-\eqref{fieldeqs-5_dmls_approx} (see also Fig.~\ref{fig_mass_gc_approx_Maxw})
that the dependence of the maximum mass on the charge is
\begin{equation}
\label{M_max_approx_with_charge}
	M_{*}^{\text{max}} \approx \beta(\bar Q_M) \sqrt{|\bar \lambda|}\frac{ M_\text{Pl}^2}{\mu},
\end{equation}
where the numerical values of the coefficient $ \beta(\bar Q_M)$ are given in Table~\ref{tab1}. This coefficient is well approximated by the formula
$\beta \approx 0.38 /\sqrt{\bar Q_{\text{crit}}-\bar Q_M}$ which determines the divergence of the maximum mass in the limit
$\bar Q_M\to\bar Q_{\text{crit}}=1$. This divergence is also illustrated in Fig.~\ref{fig_mass_gc_approx_Maxw}
by the curve joining the maxima of the mass.

\begin{table}[h]
\begin{tabular}{|c|c|c|c|c|c|c|c|c|}
\hline
$\bar Q_M$ & 0 & 0.5 & 0.7 & 0.9 & 0.925 & 0.95 & 0.975 & 0.99\\
\hline
$\beta$ & 0.41 & 0.50 & 0.64 & 1.16 & 1.35 & 1.66 & 2.38 & 3.79\\
\hline
\end{tabular}
\caption{The calculated values of the coefficient  $\beta$ from \eqref{M_max_approx_with_charge} as a function of the charge $\bar Q_M$.
}
\label{tab1}
\end{table}

Notice here that the numerical computations indicate that the above approximate solutions describe fairly well only systems located near maxima of the mass,
and the deviations from the exact solutions become stronger the further away we are from the maximum.

\subsubsection{Effective radius}
\label{eff_rad_Maxw}

In modeling ordinary stars (for example, neutron ones), it is usually assumed that they have a surface on which the pressure of matter vanishes.
In the case of field configurations (for instance, boson stars) supported by exponentially damped fields,  such a surface is already absent.
Therefore for such configurations one uses some effective radius which can be introduced in several different ways~\cite{Schunck:2003kk}.

Since the configurations considered in this section contain a long-range Maxwell field, a definition of the effective radius is different from that used in the case of configurations
consisting of exponentially damped fields. Here we follow Ref.~\cite{Jetzer:1989av}
(where charged boson stars with a Maxwell field are under consideration) and introduce the following expression for the effective radius:
\begin{equation}
R=\frac{1}{N_f}\int r j^t d^3 x=\frac{\lambda_c}{N_f}\int_0^\infty \frac{\bar f^2+\bar g^2}{\sqrt{N}}x^3 dx,
\label{eff_radius}
\end{equation}
where the particle number $N_f$ is taken from \eqref{part_num} (without the numerical coefficient before the integral).
As in the case of charged boson stars of Ref.~\cite{Jetzer:1989av}, this expression yields a finite result, in contrast to the case when the effective radius is defined in terms
of the mass integral, as is done for uncharged configurations (for details see Refs.~\cite{Schunck:2003kk,Jetzer:1989av}).

\begin{table}[h]
\begin{tabular}{|c|c|c|c|c|c|c|c|c|}
\hline
$\bar Q_M$ & 0 & 0.5 & 0.7 & 0.9 & 0.925 & 0.95 & 0.975 & 0.99\\
\hline
$\gamma_{0}$ & 2.88 & 3.39 & 3.89 & 5.68 &\ldots  &7.90  & \ldots & 17.10\\
\hline
$\gamma_{l}$ & 1.08 & 1.29 & 1.37 & 2.37 & 2.69 & 3.25 & 4.51 & 7.01\\
\hline
\end{tabular}
\caption{The calculated values of the coefficients $\gamma_{0}$ and $\gamma_{l}$ from \eqref{R_eff_lambda_0} and \eqref{R_eff_lambda_large} as a function of the charge $\bar Q_M$.
}
\label{tab2}
\end{table}

Using the expression~\eqref{eff_radius}, we have obtained the following dependencies of the effective radius on  $\bar Q_M$
for the configurations with maximum masses considered in Secs.~\ref{num_sol_Max_linear} and \ref{lim_conf_M}:
\begin{eqnarray}
	&& R^{\text{max}}=\lambda_c \gamma_{0}(\bar Q_M) \quad \quad\quad\text{for} \quad \bar\lambda=0,
\label{R_eff_lambda_0}\\
&& R_*^{\text{max}}\approx \lambda_c \gamma_{l}(\bar Q_M)\sqrt{|\bar\lambda|} \quad \text{for} \quad |\bar\lambda|\gg 1.
\label{R_eff_lambda_large}
\end{eqnarray}
The numerical values of the coefficients $\gamma_{0}$ and $\gamma_{l}$ appearing here are given in Table~\ref{tab2}.

Asymptotically, as $\bar Q_M\to \bar Q_{\text{crit}}=1$, these coefficients are approximated by the expressions
$\gamma_{0} \approx 1.7 /\sqrt{\bar Q_{\text{crit}}-\bar Q_M}$ and $\gamma_{l} \approx 0.73 /\sqrt{\bar Q_{\text{crit}}-\bar Q_M}$
which determine the divergences of the radii as the charge tends to the critical value.

\subsection{The case of the Proca field}
\label{Proca_field}

For the Proca field Eqs.~\eqref{fieldeqs-1_dmls}-\eqref{fieldeqs-5_dmls} are solved
when $\alpha\neq 0$ and with boundary conditions assigned at the center in the form of~\eqref{bound_cond}.
In doing so, solutions obtained near the center are smoothly matched with asymptotic solutions obtained as $x\to \infty$, which are
\begin{equation}
\bar f\approx \bar f_\infty e^{-\sqrt{1-\bar E^2}\,x}+\ldots, \quad \bar g\approx \bar g_\infty e^{-\sqrt{1-\bar E^2}\,x}+\ldots, \quad \bar \phi\approx \bar \phi_\infty \frac{e^{-\alpha x}}{x}+\ldots,
\quad \sigma\approx 1+\ldots, \quad \bar m\approx \bar m_\infty+\ldots,
\label{asymp_Proca}
\end{equation}
where $\bar f_\infty, \bar g_\infty, \bar \phi_\infty, \bar m_\infty$ are integration constants, and, as in the case of the Maxwell field, the constant
$\bar m_\infty$ plays the role of the total ADM mass of the configurations under consideration.

\subsubsection{Linear spinor fields}
\label{lin_spinor_Proca}

\begin{figure}[t]
\begin{minipage}[t]{.49\linewidth}
		\begin{center}
			\includegraphics[width=.98\linewidth]{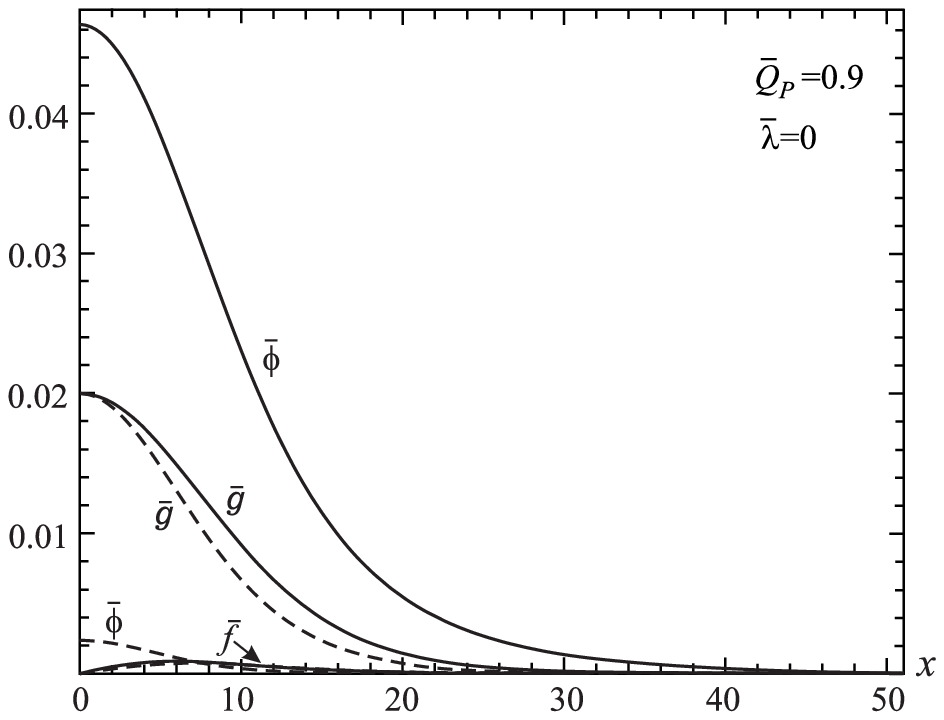}
		\end{center}
\vspace{-0.5cm}
		\caption{Proca field: the distributions of the fields $\bar g$, $\bar f$, and $\bar\phi$ along the dimensionless radius $x$.
The solid lines correspond to the solutions with  $\alpha=0.1$, the dashed lines to those with $\alpha=1$.
 		}
		\label{fig_field_distr_Proca}
	\end{minipage}
\hfill
	\begin{minipage}[t]{.49\linewidth}
		\begin{center}
			\includegraphics[width=.98\linewidth]{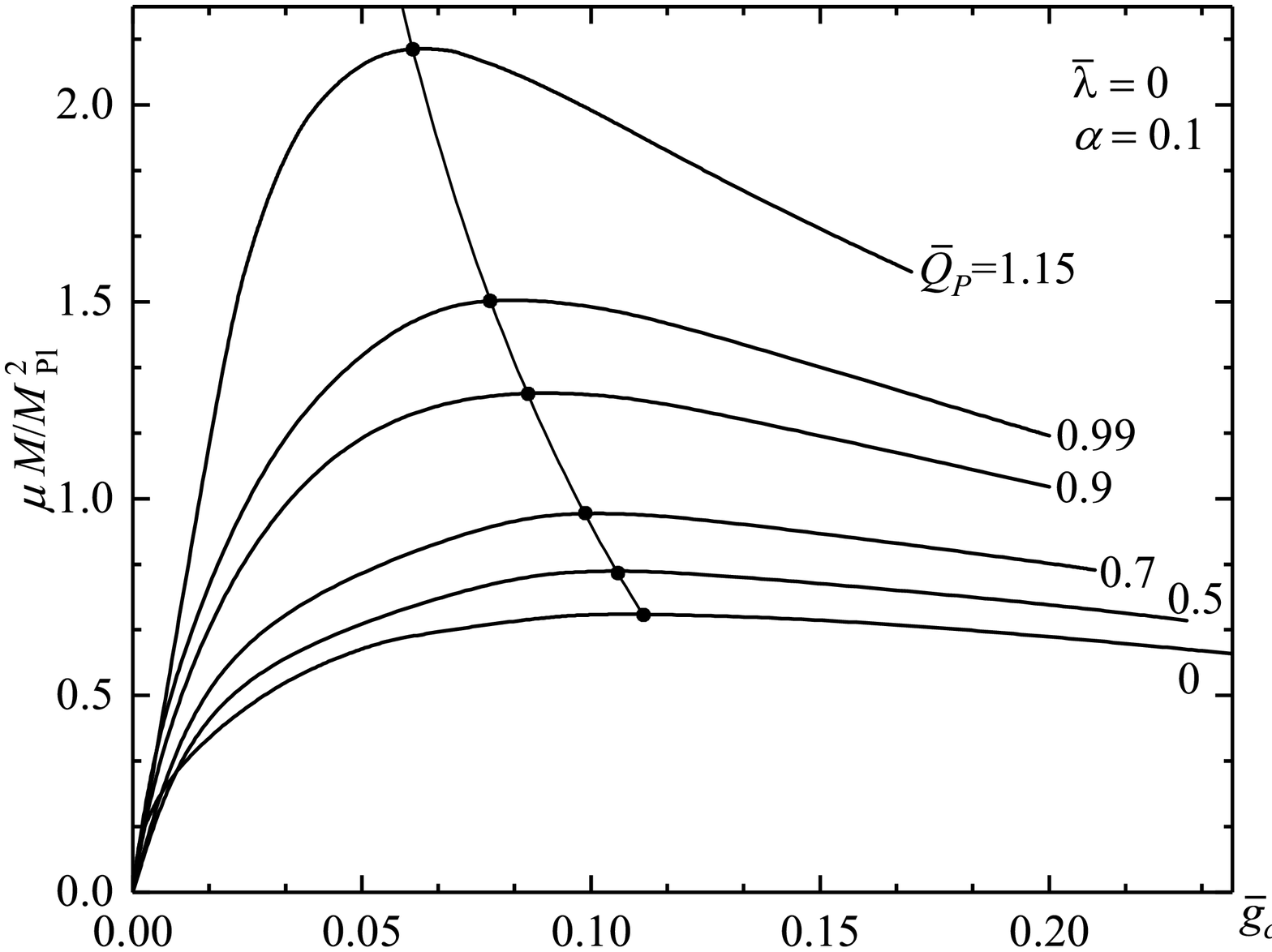}
		\end{center}
\vspace{-0.5cm}
		\caption{Proca field: the dimensionless total mass of the configurations $\bar m_\infty$ as a function of $\bar g_c$ for the systems with $\bar \lambda=0$ and
$\bar Q_P=0, \,0.5,\, 0.7,\, 0.9, 0.99, 1.15$. The bold dots show the positions of maxima of the mass.
		}
		\label{fig_M_Q_Proca}
	\end{minipage}
\end{figure}

As in the case of the massless vector field of Sec.~\ref{Maxw_field},
here we seek zero-node regular asymptotically flat solutions describing configurations with finite energies and total masses.
In doing so, in contrast to the case of the Maxwell field, besides the eigenvalue $\bar E$, it is also necessary to seek an eigenvalue of $\bar \phi_c$
ensuring exponential damping of the Proca field at infinity.

Since the Proca field is massive, it decays exponentially fast with a rate
determined by the magnitude of the parameter $\alpha$ [see Eq.~\eqref{asymp_Proca}].
When $\alpha\to 0$, we return to the exterior Coulomb-type solution~\eqref{asymp_sol}. In turn,
when $\alpha$ is sufficiently large, the field $\bar \phi$ is concentrated inside the radius $x_b$.
The typical distributions of the matter fields for the two values $\alpha=0.1$ and $\alpha=1$ and for the fixed central value $\bar g_c=0.02$
are given in Fig.~\ref{fig_field_distr_Proca}. One can see that, with increasing $\alpha$, the field $\bar \phi$
is increasingly concentrated inside the radius where the spinor fields are nonvanishing, and the central value $\bar \phi_c$ decreases in turn.

By varying $\bar Q_P$, we have obtained the dependencies of the mass of the configurations on the value of $\bar g_c$
shown in Fig.~\ref{fig_M_Q_Proca}. As in the case of the Maxwell field, in plotting these curves we have kept track of the sign of the binding energy using the expression
\eqref{part_num}. Correspondingly, the dependencies in Fig.~\ref{fig_M_Q_Proca} are given only for positive values of the BE.
It is seen from the results obtained that, with increasing $\bar Q_P$, the maximum mass of the systems under consideration increases, shifting towards smaller values of
$\bar g_c$. In turn, from the behavior of the curve joining the maxima of the mass, one sees the growth of the mass with increasing $\bar Q_P$.
However, in contrast to the case of the Maxwell field, here, apparently, there is no 
{\it finite} value of the coupling constant $\bar Q_P$ for which the mass tends to infinity, but it increases monotonically
as $\bar Q_P\to \infty$ (regarding this issue, see also the next paragraph).

\subsubsection{The case of $|\bar \lambda| \gg 1$}

\begin{figure}[t]
\begin{minipage}[t]{.49\linewidth}
		\begin{center}
			\includegraphics[width=1\linewidth]{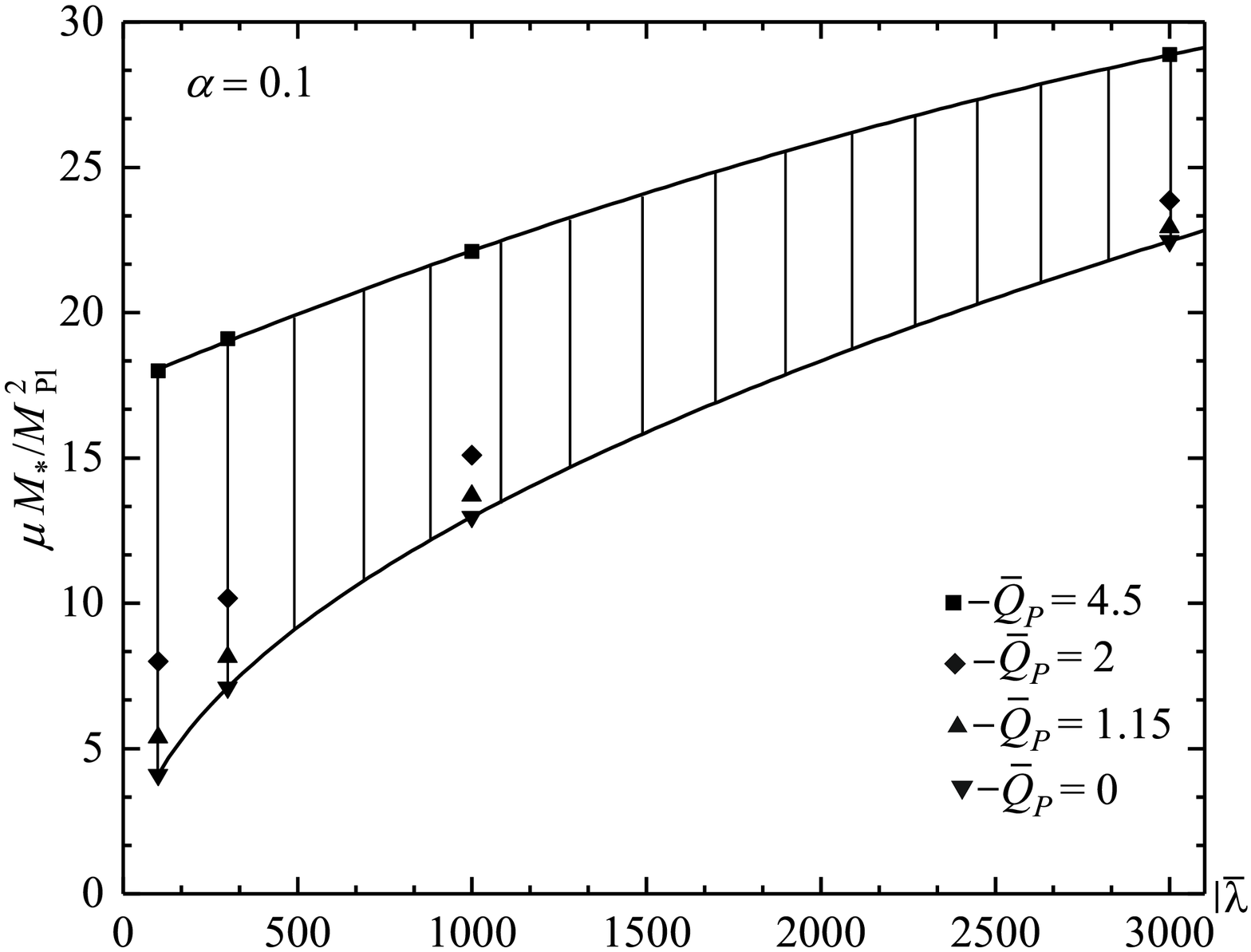}
		\end{center}
\vspace{-0.5cm}
		\caption{Proca field: the symbols show the calculated values of
the maximum masses of the configurations with $|\bar \lambda| \gg 1$
for several values of the coupling constant $\bar Q_P$. Masses of the configurations with intermediate values of
$\bar Q_P$ and $|\bar \lambda|$ lie within the shaded region between the two interpolation curves.
		}
		\label{fig_mass_lambda_Proca}
	\end{minipage}
\hfill
	\begin{minipage}[t]{.49\linewidth}
		\begin{center}
			\includegraphics[width=1\linewidth]{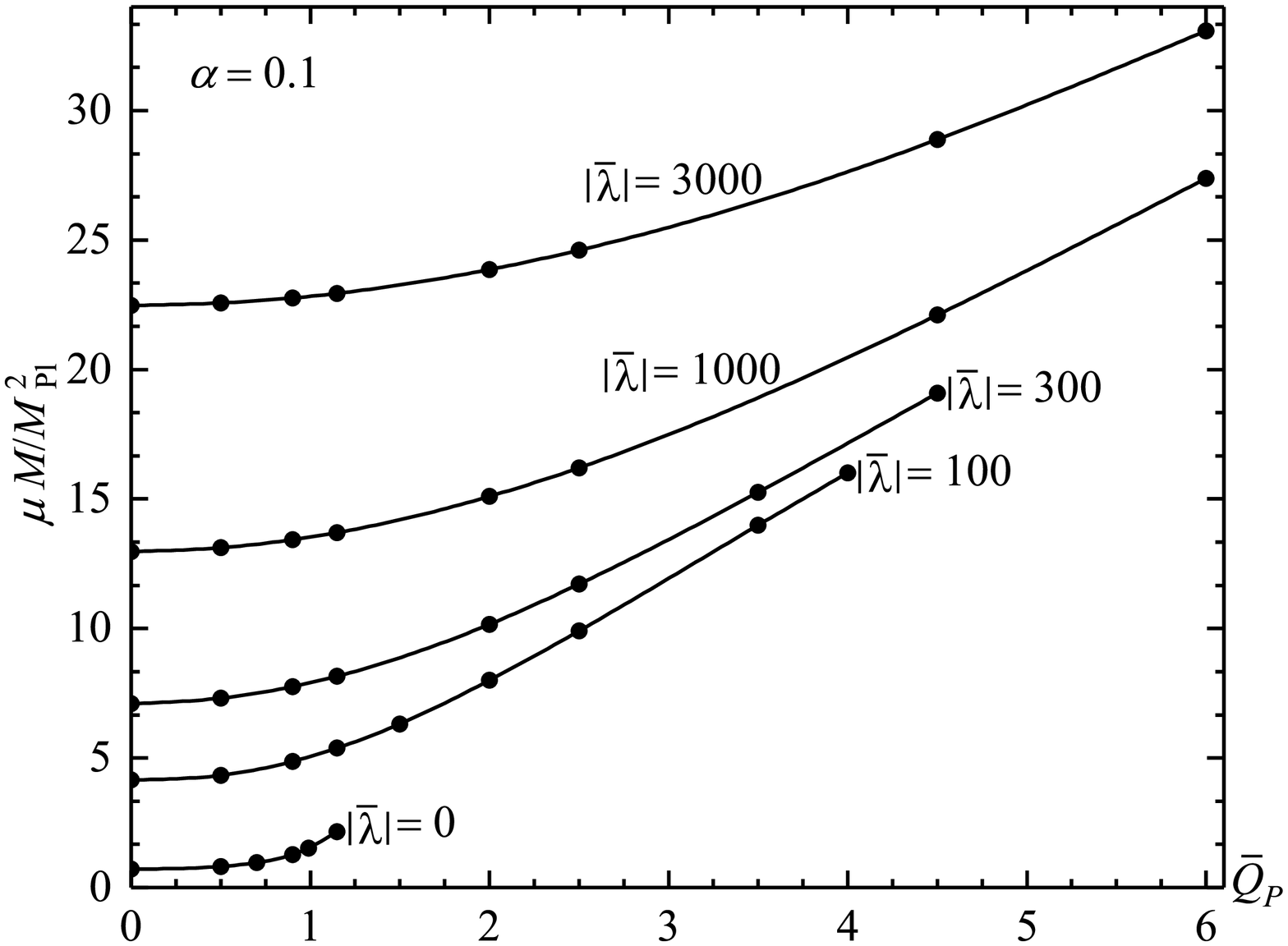}
		\end{center}
\vspace{-0.5cm}
		\caption{Proca field: the maximum masses of the configurations
{\it vs.} the coupling constant $\bar Q_P$ for different values of $|\bar \lambda|$. The rightmost points of the curves correspond to the values of
$\bar Q_P$ to which we have succeeded (technically) in obtaining the solutions.
		}
		\label{fig_mass_Q_Proca}
	\end{minipage}
\end{figure}

Consider now the case of negative $|\bar \lambda| \gg 1$. Numerical computations indicate that, as for the problem of Sec.~\ref{lim_conf_M},
in this case the spinor field $\bar f$  is also much smaller than $\bar g$. Then the leading term in Eq.~\eqref{fieldeqs-1_dmls} is again
the third term $(\ldots)\bar g$, and this equation yields (in the approximation of $\bar f \ll \bar g$)
\begin{equation}
 \label{g_approx_Proca}
 \bar g^2 \approx \frac{1}{8\bar \lambda}
 \left[1 - \frac{1}{\sigma\sqrt{N}}\left(\bar E-\bar Q_P\bar\phi\right)\right].
\end{equation}

Unfortunately, on account of  the presence of the massive vector field, in this case
the field equations are not scale invariant, and therefore it is impossible to introduce the mass and the radius rescaled through
$|\bar \lambda|$, as is done in Sec.~\ref{lim_conf_M}. Nevertheless, one can use the approximate expression
 \eqref{g_approx_Proca}, insertion of which into Eqs.~\eqref{fieldeqs-3_dmls}-\eqref{fieldeqs-5_dmls}
 yields an approximate set of equations for the metric functions and the Proca field. Solving this set of equations numerically, one can show that,
 as in the case of the Maxwell field, the approximate solution~\eqref{g_approx_Proca} agrees well with the exact solution, except only the region at large radii
 (cf. Fig.~\ref{fig_field_distr}). This permits us to use approximate set of equations thus derived to describe configurations with the Proca field.
 In this case there is only one eigenparameter ($\bar E$ or $\bar \phi_c$) whose value can be found by using the shooting method;
 this is technically much easier since to obtain a solution to the exact equations it would be necessary to adjust both aforementioned parameters.

Numerically solving Eqs.~\eqref{fieldeqs-3_dmls}-\eqref{fieldeqs-5_dmls} in the approximation~\eqref{g_approx_Proca},
we have plotted in Fig.~\ref{fig_mass_lambda_Proca} the dependencies of the maximum masses of the configurations on $|\bar\lambda|$ for the one value of the parameter
$\alpha=0.1$ and for different $\bar Q_P$. In this figure, the bottom curve corresponds to the interpolation formula
 $	M_{*}^{\text{max}}\approx 0.41 \sqrt{|\bar \lambda|}M_{\text{Pl}}^2/\mu$
 describing the dependence of the mass on $|\bar \lambda|$ for the systems without a vector field ($\bar Q_P=0$).
 For the systems with the vector field ($\bar Q_P\neq0$), the masses are not already proportional to $\sqrt{|\bar \lambda|}$;
 this is demonstrated by the top interpolation curve plotted for the case of
 $\bar Q_P=4.5$. Hence we see that, in contrast to the case of the Maxwell field where for any value of the coupling constant $\bar Q_M$
 the mass is proportional to $\sqrt{|\bar \lambda|}$
[see Eq.~\eqref{M_max_approx_with_charge}], in the case of the Proca field, this is not so.
It is evident that this is due to the scale noninvariance of the Proca field.

It is seen from Fig.~\ref{fig_mass_lambda_Proca} that for any fixed value of $\bar \lambda$ the maximum mass of the configurations grows with increasing
$\bar Q_P$. This is illustrated in more detail in Fig.~\ref{fig_mass_Q_Proca}, which shows the dependencies of the mass on
 $\bar Q_P$ for different $\bar \lambda$. From an analysis of the behavior of these curves,
 one would expect that asymptotically (for $\bar Q_P\gg 1$) they will converge to one curve, and the mass in turn will diverge as $\bar Q_P\to \infty$.
However, this question requires more detailed study, including a consideration of the influence that the magnitude of the parameter $\alpha$ has on these dependencies.

\subsubsection{Effective radius}

\begin{figure}[t]
		\begin{center}
			\includegraphics[width=.5\linewidth]{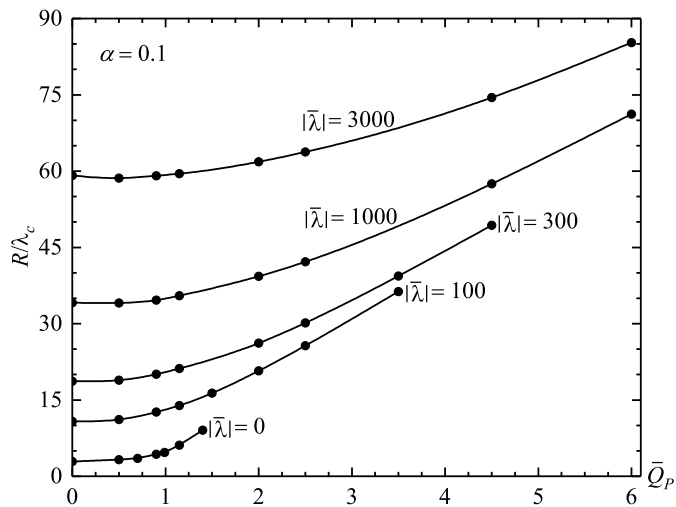}
		\end{center}
\vspace{-0.5cm}
		\caption{Proca field: the effective radii of the configurations with
maximum masses {\it vs.} the coupling constant $\bar Q_P$ for different values of $|\bar \lambda|$.
		}
		\label{fig_mass_R_Proca}
\end{figure}

Since the spinor and Proca fields  under consideration decrease exponentially fast with distance  [see Eq.~\eqref{asymp_Proca}],
to define the effective radius, we may employ one of the ways applied for boson stars.
Namely, we choose the following definition which is introduced using the
mass integral (see, e.g., in Ref.~\cite{Schunck:2003kk}):
$$
	R = \frac{\int_0^\infty T^0_0 r^3 dr }{\int_0^\infty T^0_0 r^2 dr} =
	\lambda_c
	\frac{\int_0^\infty \bar T^0_0 x^3 dx}{\int_0^\infty \bar T^0_0 x^2 dx } ,
$$
where the energy density $T^0_0$ is taken from Eq.~\eqref{EM} or, already in the dimensionless form, from the right-hand side of Eq.~\eqref{fieldeqs-3_dmls}.
Using this expression, we have plotted in Fig.~\ref{fig_mass_R_Proca} the dependencies of the effective radii of the configurations with maximum masses on the coupling constant
$\bar Q_P$ for different $\bar \lambda$. One sees from this figure that, as in the case of the masses
(see Fig.~\ref{fig_mass_Q_Proca}), there is a monotonic growth of the radii with increasing $\bar Q_P$. Asymptotically
(for $\bar Q_P\gg 1$), one might expect that all the curves will converge to one curve which will determine the divergence of the radii as $\bar Q_P\to \infty$.

\section{Conclusions and discussion}
\label{concl}

We have studied compact strongly gravitating configurations consisting of nonlinear spinor fields minimally coupled to vector (Maxwell and Proca) fields.
In order to ensure spherical symmetry of the system, we have used two spinor fields having opposite spins, and this
enabled us to get a diagonal energy-momentum tensor.
Consistent with this, we have found families of equilibrium configurations described by regular zero-node asymptotically flat solutions for
the static vector fields and for the stationary spinor fields, oscillating with a frequency $E/\hbar$. It was shown that for all values of $E$ and of the  coupling constants $\lambda, Q_{M,P}$
which we have considered, these solutions describe configurations possessing a positive ADM mass. This enables one to apply such solutions in modeling compact gravitating objects (Dirac stars).

The main purpose of the paper was to examine the influence of the presence of the Maxwell and Proca fields on the physical
characteristics of the Dirac stars. As in the case of the Dirac stars supported only by nonlinear spinor fields~\cite{Dzhunushaliev:2018jhj},
our goal here was to explore the possibility of obtaining objects with masses of the order of the Chandrasekhar mass.
For this purpose, we have studied in detail the cases with large negative values of the dimensionless coupling constant $\bar \lambda$ and with different values of the dimensionless coupling constants $\bar Q_{M,P}$.
The results obtained can be summarized as follows:
\begin{enumerate}
\itemsep=-0.2pt
\item[(I)] For the case of the Maxwell field, we have considered all physically admissible values of the charge $0 \leq \bar Q_M < 1$.
In this case the families of  equilibrium configurations can be parametrized by two dimensionless quantities:
 the coupling constant $\bar \lambda =\lambda M_\text{Pl}^2 c/4\pi \hbar^3$ and the charge $\bar Q_M$. Due to the
 scale invariance of the system, in the limit  $|\bar \lambda|\gg 1$, it is possible to get approximate equations which do not contain
   $\bar \lambda$ explicitly.
Consistent with the dimensions of  $[\lambda]=\text{erg cm}^3$, one can assume that its characteristic value is
  $\lambda \sim  \tilde \lambda \,\mu c^2 \lambda_c^3$, where the dimensionless quantity $\tilde \lambda \sim 1$.
 Then the dependence of the maximum mass of the systems in question on
  $|\bar \lambda|$  and $\bar Q_M$ [see Eq.~\eqref{M_max_approx_with_charge}] can be represented as
$$
	M_*^{\text{max}}\approx \beta(\bar Q_M) \sqrt{|\bar \lambda|}\frac{M_\text{Pl}^2}{\mu} \approx 0.46\, \beta(\bar Q_M)\sqrt{|\tilde \lambda|}M_{\odot}\left(\frac{\text{GeV}}{\mu}\right)^2,
$$
where the numerical values of the coefficient $\beta(\bar Q_M)$ are taken from Table~\ref{tab1}, and they are well approximated by the formula
 $\beta \approx 0.38 /\sqrt{\bar Q_{\text{crit}}-\bar Q_M}$ which determines the divergence of the maximum mass in the limit
$\bar Q_M\to\bar Q_{\text{crit}}=1$. The above mass $M_*^{\text{max}}$
is comparable to the Chandrasekhar mass for the typical mass of a fermion $\mu\sim 1~\text{GeV}$.
In this respect the behavior of the dependence of the maximum mass of the charged Dirac stars considered here  on the coupling constant $\bar \lambda$ and on the charge $\bar Q_M$
 is similar to that of charged boson stars of Ref.~\cite{Jetzer:1989av}.

In turn, the dependence of the effective radii of the limiting configurations with maximum masses on
    $|\bar \lambda|$  and $\bar Q_M$  [see Eq.~\eqref{R_eff_lambda_large}] can be represented as
$$
R_*^{\text{max}}\approx \lambda_c \gamma_{l}(\bar Q_M)\sqrt{|\bar\lambda|}\approx
0.68\, \gamma_{l}(\bar Q_M)\sqrt{|\tilde \lambda|}\left(\frac{\text{GeV}}{\mu}\right)^2 \,\, \text{km},
$$
 where the numerical values of the coefficient $\gamma_{l}(\bar Q_M)$ are taken from Table~\ref{tab2}, and in the limit
 $\bar Q_M\to\bar Q_{\text{crit}}=1$ they are approximated by the formula
$\gamma_{l} \approx 0.73 /\sqrt{\bar Q_{\text{crit}}-\bar Q_M}$ which determines the divergence of the radii when the charge tends to the critical value.
For the typical mass of a fermion $\mu\sim 1~\text{GeV}$, the above expression gives the radii of the order of kilometers. In combination  with the masses
of the order of the Chandrasekhar mass (see above), this corresponds to characteristics typical for neutron stars.

 \item[(II)]  In the case of the Proca field, besides the coupling constants $\bar \lambda$ and $\bar Q_P$, the system involves one more free parameter $\alpha$
 equal to the ratio of the Proca mass to the mass of the spinor field. In the limit $\alpha\to 0$, we return to the results of item (I).
 When $\alpha\neq 0$, the vector field is not already a long-range one, and it decreases exponentially fast with distance according to the asymptotic law given by Eq.~\eqref{asymp_Proca}.

Due to the absence of the scale invariance, in this case it is already impossible to get rid of $\bar \lambda$ in the approximate equations valid for $|\bar \lambda|\gg 1$,
and therefore the only possibility is to obtain solutions for particular values of~$\bar \lambda$.
Numerical calculations indicate that in the case of the Proca field,
the maximum mass ceases to alter proportionally to $\sqrt{|\bar \lambda|}$, and, with increasing $|\bar \lambda|$, it increases with a rate which
is in general determined by the values of $\bar Q_P$ and $\alpha$ (see Fig.~\ref{fig_mass_lambda_Proca}).

 In turn, consistent with the numerical results obtained, one may assume that the system seems not to involve a finite value of the coupling constant $\bar Q_P$
 for which the total mass and radius of the configurations would diverge. With increasing
   $\bar Q_P$, a monotonic increase in the maximum masses and in the corresponding effective radii takes place
   (see Figs.~\ref{fig_mass_Q_Proca} and~\ref{fig_mass_R_Proca}), and one might naively expect that they will diverge only as $\bar Q_P\to \infty$.
  But this question requires further investigation.
\end{enumerate}

We can draw from the above results that the physical characteristics of the configurations under investigation are largely determined by the values of the coupling constants
 $\lambda, Q_{M,P}$ and by the ratio of the masses $\alpha$. Since in the present paper we have studied only a finite set of values of the above parameters
 for the system with the Proca field, it seems to be of interest to extend the range of values of these parameters by considering, in particular, large values of $\alpha$ and~$Q_{P}$.

In conclusion, let us briefly address the question of stability of the systems under investigation. As is seen from the above results, all the configurations considered here
 can be parametrized by the central value of the spinor field~$g_c$. The total mass is then a function of this parameter,
and for any values of the coupling constants $\lambda,Q_{M,P}$ and of the parameter $\alpha$, there exists a first peak in the mass (a local maximum).
Similarly to models of neutron and boson stars, one can naively expect that a transition through this local maximum should lead to instability against perturbations which compress the entire
star as a whole. However, this question requires special consideration by analyzing the stability of
the configurations studied here against, for instance, linear perturbations, as is done for boson stars~\cite{Gleiser:1988ih,Jetzer:1989us},
or by using catastrophe theory~\cite{Kusmartsev:1990cr}.

\section*{Acknowledgments}
The authors gratefully acknowledge support provided by Grant No.~BR05236494
in Fundamental Research in Natural Sciences by the Ministry of Education and Science of the Republic of Kazakhstan. We are grateful to the Research Group
Linkage Programme of the Alexander von Humboldt Foundation for the support of this research.

\end{document}